\documentclass[aps,prb,reprint,superscriptaddress,showpacs,amsmath,amssymb]{revtex4-2}

\usepackage{graphicx, comment}
\usepackage{dcolumn}
\usepackage{bm}
\usepackage{textcomp}
\usepackage{lineno}

\begin{document}


\title{Multiresonances of quasi-trapped modes in metasurfaces based on nanoparticles of transition metal dichalcogenides}


\author{M.Yu. Gubin}
\affiliation{Department of Physics and Applied Mathematics, Vladimir State University named after Alexander and Nikolay Stoletovs (VlSU), Vladimir 600000, Russia}
\affiliation{Center for Photonics and 2D Materials, Moscow Institute of Physics and Technology (MIPT), Dolgoprudny 141701, Russia}
\author{A.V. Shesterikov}
\affiliation{Department of Physics and Applied Mathematics, Vladimir State University named after Alexander and Nikolay Stoletovs (VlSU), Vladimir 600000, Russia}
\affiliation{Center for Photonics and 2D Materials, Moscow Institute of Physics and Technology (MIPT), Dolgoprudny 141701, Russia}
\author{V.S. Volkov}
\affiliation{Center for Photonics and 2D Materials, Moscow Institute of Physics and Technology (MIPT), Dolgoprudny 141701, Russia}
\author{A.V. Prokhorov}
\email{alprokhorov33@gmail.com}
\affiliation{Department of Physics and Applied Mathematics, Vladimir State University named after Alexander and Nikolay Stoletovs (VlSU), Vladimir 600000, Russia}
\affiliation{Center for Photonics and 2D Materials, Moscow Institute of Physics and Technology (MIPT), Dolgoprudny 141701, Russia}



\begin{abstract}
The features of polarization control of multiple multiresonances for quasi-trapped modes excited by synchronization of bianisotropic dipole responses in MoS$_2$ disks with a hole are considered. Using the numerical calculations with analytical multipole analysis, we showed that the presence of a strong optical anisotropy of MoS$_2$ nanoparticles provides an additional degree of freedom and allows to observe several resonances of electric and magnetic dipoles at once in a narrow spectral range. Based on the simulation results, we obtained the frequency dependences for the dipole polarizabilities of the MoS$_2$ disk with a hole, which allow one to distinguish the contributions of the nonlocal and bianisotropic dipole responses and possessing several features in the near infrared range. Using the polarizability spectra of single nanoparticles and applying the tuning strategy, the design of the MoS$_2$ metasurface supporting three resonances of quasi-trapped modes at once in a narrow spectral range was developed. One of these resonances corresponds to the telecom wavelength of 1550~nm. The spectrum of light reflection for MoS$_2$ metasurface is characterized by three narrowband dips corresponding to the wavelengths of the quasi-trapped modes. It was shown that a change in the polarization of a wave normally incident on the metasurface to orthogonal one leads to a change in the type of bianisotropic response excited in each MoS$_2$ disk and to the excitation of three other features in the reflection spectra of the metasurface at wavelengths close to the initial values.
\end{abstract}


\maketitle

\section{Introduction}
In general, the trapped modes or TMs are protected eigenmodes of ideal lossless optical systems, which are characterized by high quality factors~\cite{t16,Koshelev} and detected via narrow features in the reflection/transmission spectra. Such states can be obtained by symmetry breaking for either individual building blocks or the entire lattice~\cite{Fedotov,Tuz2018}. In particular, nanoparticles (NPs) with broken symmetry of the shape can support a bianisotropic response, for which a pump wave even normally incident to the nanoparticle’s base can excite dipole components in the nanoparticle that are longitudinal to its wave vector~\cite{Evlyukhin2020}. By choosing the period of lattice composed of such bianisotropic building blocks, it is possible to synchronize the phases of the dipoles of individual NPs, which results in the formation of TM in the metasurface. At the same time, introducing the defects into the metasurface’s structure leads to the opening of a channel for energy leakage from TM and, hence, to the decrease of its quality factor, and such a mode is called a quasi-trapped mode (QTM).

In the case when the building blocks of the metasurface are made of high-anisotropic materials, an additional degree of freedom arises allowing one to control the spectral position of the QTM, as well as its characteristics~\cite{ourPRB2022}. The bulk transition metal dichalcogenides (TMDCs) are good candidates to be used as such materials, because they possess a unique combination of high refractive index~\cite{Verre} and strong optical anisotropy~\cite{Ermolaev2021}. The first feature allows to observe the bright Mie resonances in the visible and near infrared ranges for nanoparticles fabricated from such a material~\cite{Tselikov2022,Verre}. The second one determines the large birefringence, which can be used for creation of ultra-thin polarizing devices~\cite{LinTseng}.
\begin{figure*}[t]
\centering
\includegraphics[width=1.7\columnwidth]{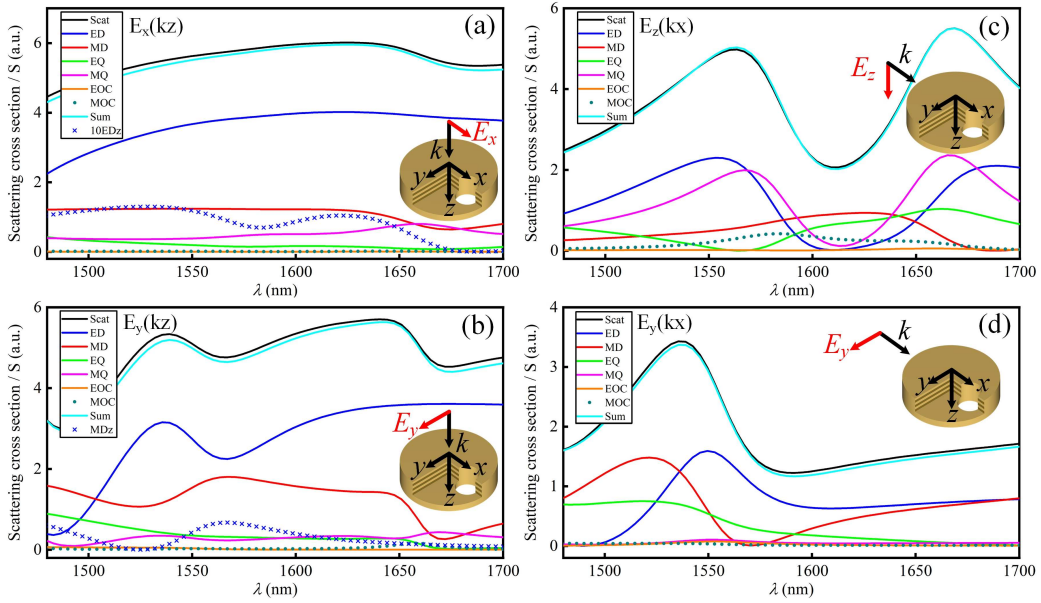}
\caption{Spectra of the scattering cross sections with corresponding multipole decomposition calculated for the MoS$_2$ disk (the radius and height $R_2=H=418$~nm) with the through hole shifted from the disk's center on the distance $\Delta_x=159$~nm with radius $R_1=133$~nm and for different irradiation conditions clarified by the insets. Orientation of the MoS$_2$ layers is parallel to the $xy$ plane.}\label{fig1}
\end{figure*}

One of the remarkable features of QTMs in metasurfaces is their polarization sensitivity, when change in the polarization of the exciting field to the orthogonal one can lead, in particular, to switching between electric and magnetic type of QTM~\cite{t16}. This can be used, among other applications, to design selective polarizing mirrors and optical router~\cite{GuoDecker,YanYang,KerenZur,HuWang}. At the same time, to increase the coding density and bandwidth of such devices, it is expedient to realize the conditions corresponding to the excitation of several QTMs at once in the metasurface and switching between them.

In this paper, we propose a strategy for design of all-dielectric metasurfaces supporting several quasi-trapped modes at close frequencies at once. Such a possibility is achieved through the use of MoS$_2$ as a material of the building blocks of the metasurface. In particular, MoS$_2$ disks possessing strong optical anisotropy can have several dipole resonances at once in a narrow range of the infrared region under lateral irradiation. This leads to the appearance of a set of features for the spectral dependences of their electric and magnetic polarizability. Using these features and optimizing the parameters of the metasurface composed of MoS$_2$ disks with a shifted hole, one can obtain the conditions for excitation of three QTM resonances in metasurface at once caused by electric bianisotropy in a narrow spectral range. A distinctive feature of the designed metasurfaces is the possibility of the polarization switching between the presented resonances and other multiresonances of the QTM associated with the anisotropic magnetic-type response in disks. In the case of metasurface irradiated by a wave with polarization oriented at an angle to the direction of the hole shift, each peak in the reflection spectra of the metasurface can be transformed into a double feature, which is caused by the asymmetry of the disk shape. In addition, the presence of a strong anisotropy of MoS$_2$ material can lead to the cross-polarization effect accompanied by energy transfer into the polarization component of the reflected/transmitted wave that is orthogonal to the polarization of the incident wave.

\section{Features of the polarizability of a single MoS$_2$ disk with a defect}
We choose a building block of a metasurface in the form of a disk fabricated from anisotropic MoS$_2$ material~\cite{Ermolaev2021} with layers arranged parallel to the disk's base. The radius and height of the disk are $R_2=H=418$~nm and disk has a round penetrating hole shifted from the disk's center along the x axis on the distance $\Delta_x=159$~nm. Numerical calculation of the scattering spectra of NPs was carried out using COMSOL Multiphysics. Multipole decomposition was calculated on the basis of our own algorithms realized in the same software. As a result, under normal irradiation of such a disk by waves $E_x(k_z)$ in Figure~\ref{fig1}a and $E_y(k_z)$ in Figure~\ref{fig1}b, the scattering cross section has a smooth dependence without any features in the wavelength range 1480--1700~nm. However, there is an excitation of minor components of the electric dipole $\textup{ED}_z$ in Figure~\ref{fig1}a and magnetic dipole $\textup{MD}_z$ in Figure~\ref{fig1}b oriented perpendicular to the disk's base.

In addition, under lateral irradiation of the same disk by waves $E_z(k_x)$ in Figure~\ref{fig1}c and $E_y(k_x)$ in Figure~\ref{fig1}d, we find several features at once in its scattering spectrum. To analyze the spectra of the scattering cross section, we use the multipole decomposition in the form:
\begin{eqnarray}
\label{eq1}
\nonumber
\sigma_{scat}&\cong&\frac{k^4}{6\pi\varepsilon_0^2|\textbf{E}|^2}|\textbf{p}|^2+\frac{k^4\varepsilon_d\mu_0}{6\pi\varepsilon_0|\textbf{E}|^2}|\textbf{m}|^2\\
\nonumber
&+&\frac{k^6\varepsilon_d}{720\pi\varepsilon_0^2|\textbf{E}|^2}\sum_{\alpha\beta} |Q_{\alpha\beta}|^2+\frac{k^6\varepsilon_d^2\mu_0}{80\pi\varepsilon_0|\textbf{E}|^2}\sum_{\alpha\beta} |M_{\alpha\beta}|^2\\
&+&\frac{k^8\varepsilon_d^2}{1890\pi\varepsilon_0^2|\textbf{E}|^2}\sum_{\alpha\beta\gamma} |O_{\alpha\beta\gamma}|^2,
\end{eqnarray}
where $k$ is the vacuum wave number, $\varepsilon_0$ and $\varepsilon_d$ are the vacuum dielectric and relative dielectric constants of surrounding medium, respectively, $\mu_0$ is the vacuum permeability, \textbf{E} is the vector of electric amplitude of the incident plane waves, \textbf{p} and \textbf{m} are the vectors of electric and magnetic dipole moments of the scatterer, respectively, and $\hat{Q}$, $\hat{M}$, and $\hat{O}$ are the tensors of electric and magnetic quadrupole moments, and electric octupole moment, respectively. Corresponding expressions for the multipole moments defined by the density of the displacement current can be found elsewhere~\cite{Evlyukhin2019}. In this article we consider that all multipole moments of individual nanoparticles are located at their center of mass. This also concerns the calculation of the dipole polarizabilities presented further.

The contributions of different multipoles (see Equation~\ref{eq1}) to the total scattering cross section shown in Figure~\ref{fig1} can be written as the sum of their spatial components in the form:
\begin{subequations}
\label{eq2}
\begin{align}
\textup{ED}&=\sum_{i} \textup{ED}_i=\frac{k^4}{6\pi\varepsilon_0^2|\textbf{\textrm{E}}|^2}\sum_{i} |p_{i}|^2,\\
\textup{MD}&=\sum_{i} \textup{MD}_i=\frac{k^4\varepsilon_d\mu_0}{6\pi\varepsilon_0|\textbf{\textrm{E}}|^2}\sum_{i} |m_{i}|^2,\\
\textup{EQ}&=\frac{k^6\varepsilon_d}{720\pi\varepsilon_0^2|\textbf{\textrm{E}}|^2}\sum_{i,j} |Qe_{ij}|^2,\\
\textup{MQ}&=\frac{k^6\varepsilon_d^2\mu_0}{80\pi\varepsilon_0|\textbf{\textrm{E}}|^2}\sum_{i,j} |Qm_{ij}|^2,
\end{align}
\end{subequations}
where $i,j=x,y,z$.

In general case, the excitation of minor components of the electric $p_z^{(E,\pm{H},\pm{k})}$ and magnetic $m_z^{(\mp{H},E,\pm{k})}$ dipole moments of NP irradiated by plane waves $E=(E_x,0,0)\exp(\pm{ikz})$ or $E=(0,E_y,0)\exp(\pm{ikz})$, respectively, can be represented on the basis of theory of nonlocal response as follows~\cite{Achouri}:
\begin{subequations}
\label{eq3}
\begin{align}
p_z^{(E,\pm{H},\pm{k})}&=\alpha_{zx}^{ee}E_x\pm a_{zxz}ikE_x\pm\alpha_{zy}^{em}H_y,\\
m_z^{(\mp{H},E,\pm{k})}&=\mp\alpha_{zx}^{mm}H_x\pm c_{zyz}ikE_y+\alpha_{zy}^{me}E_y,
\end{align}
\end{subequations}
where $\alpha_{zx}^{ee}$ ($\alpha_{zx}^{mm}$) corresponds to the direct (local) excitation of dipole moments by the electric (magnetic) field of the incident wave; $a_{zxz}$  ($c_{zyz}$) is the nonlocal response of NP, and $\alpha_{zy}^{em}$ ($\alpha_{zy}^{me}$) corresponds to bianisotropic response~\cite{t16}. The latter effect can occur in the systems with broken rotational symmetry of the particle. As a result, the nonzero longitudinal components of the dipole moments can be excited in the considered disk with a hole, which are parallel to the wave vector of the incident wave~\cite{ourPRB2022}. At the same time, the type of bianisotropic response depends on the orientation of the defect relative to the components of the electric and magnetic fields of the incident wave. In particular, if the direction between the defect and center of symmetry of the particle coincides with the electric field strength vector of the incident wave, then the longitudinal component of the electric dipole is excited in the particle, see Figure~\ref{fig1}a for $\textup{ED}_z$, and if this direction coincides with the magnetic field strength vector, then the component of the magnetic dipole is excited, see Figure~\ref{fig1}b for $\textup{MD}_z$. Switching of the type of bianisotropy excited in a single MoS$_2$ disk can be controlled by changing the polarization plane of the incident wave. At the same time, we used in our calculations the permittivity of the MoS$_2$ material measured in Refs.~\cite{Ermolaev2021,Ermolaev2020}. We also note that the appearance of the repetitive features of the scattering cross section in Figure~\ref{fig1}c and Figure~\ref{fig1}d is associated with the excitation of several resonances of the electric (ED) and magnetic (MD) dipoles at once in the considered spectral range. Therefore, one can expect the appearance of several singularity points for the polarizability at once in this range, corresponding to the intrinsic resonances of the disk.

Thus, the excitation of electric or magnetic bianisotropy of the longitudinal type in a single disk determines the possibility of creating metasurfaces based on them, supporting various types of quasi-trapped modes with the possibility of switching between them~\cite{t16}. On the other hand, the simultaneous excitation of several resonances of a single nanoparticle in a small spectral range opens up the possibility of simultaneous excitation of several lattice eigenmodes at once and the use of polarization switching between them. On the other hand, the simultaneous excitation of several resonances of a single NP in a narrow spectral range opens up the possibility of simultaneous excitation of several lattice’s eigenmodes at once and the use of polarizing switching between them.

\section{Practical calculation of the polarizabilities of a single MoS$_2$ disk with a defect}
Since the properties of metasurface are determined by the properties of individual particles, as well as the collective resonances of the entire structure, to calculate the eigenmodes of a metasurface, one should start with the calculation of the polarizabilities of its building blocks. In particular, in order to determine the conditions for collective resonance in a metasurface, it is necessary to determine the polarizability of a building block under its lateral irradiation. Since, in the considered case, we are talking about the metasurface supporting QTMs excited due to bianisotropy of both electric and magnetic types, we consider the following set of equations:
\begin{subequations}
\label{eq4}
\begin{align}
p_z^{(\pm{k},\mp{H},E)}&=\alpha_{zz}^{ee}E_z\pm a_{zzx}ikE_z\mp\alpha_{zy}^{em}H_y,\\
m_z^{(\pm{k},E,\pm{H},)}&=\pm\alpha_{zz}^{mm}H_z\pm c_{zyx}ikE_y+\alpha_{zy}^{me}E_y,\\
m_z^{(E,\pm{k},\mp{H},)}&=\mp\alpha_{zz}^{mm}H_z\pm c_{zxy}ikE_x+\alpha_{zx}^{me}E_x.
\end{align}
\end{subequations}

The Equation~(\ref{eq4}a) corresponds to the irradiation of NP in two opposite directions: forward and backward to the $x$ axis, while maintaining the direction of the vector \textbf{E} along the $z$ axis. This is sufficient to extract from Equation~(\ref{eq4}a) the expression for electric polarizability of an individual disk in the following form:
\begin{equation}
\label{eq5}
\alpha_{zz}^{ee}=\frac{p_z^{(k,-H,E)}+p_z^{(-k,H,E)}}{2E\varepsilon_0}.
\end{equation}
To determine the frequency dependences of the polarizability (\ref{eq5}), we used the results of numerical calculations for $p_z^{(k,-H,E)}$ and $p_z^{(-k,H,E)}$ when the disk is irradiated from two opposite directions. At the same time, the parameters used in the work resulted from the optimization, for which one of the disk's resonances, satisfying the condition $\textrm{Re}(1/\alpha_{zz}^{ee})=0$, corresponds to the telecom wavelength of 1550~nm. For deriving the expression (\ref{eq5}), we also used the fact that terms with $a_{zzx}$  and $\alpha_{zy}^{em}$ in Equation~(\ref{eq4}a) have opposite signs and can be canceled simultaneously. However, this is not sufficient to obtain the direct magnetic polarizability. Therefore, its values, as well as the values for the coefficient of nonlocal response $c_{zxy}$ are extracted by solving together the Equations (\ref{eq4}b) and (\ref{eq4}c) and the expressions for them take the form~\cite{ourPRB2022}:
\begin{subequations}
\label{eq6}
\begin{align}
\nonumber
\alpha_{zz}^{mm}&=\frac{1}{4H_z}\big[\big(m_z^{\left(k,E,H\right)}-m_z^{\left(-k,E,-H\right)}\big)\\
&\mspace{60mu}+\big(m_z^{\left(E,-k,H\right)}-m_z^{\left(E,k,-H\right)}\big)\big],\\
\nonumber
c_{zxy}&=\frac{1}{2ik(E_x+E_y)}\big[\big(m_z^{\left(k,E,H\right)}-m_z^{\left(-k,E,-H\right)}\big)\\
&\mspace{120mu}-\big(m_z^{\left(E,-k,H\right)}-m_z^{\left(E,k,-H\right)}\big)\big].
\end{align}
\end{subequations}

\begin{figure}[t]
\centering
\includegraphics[width=\columnwidth]{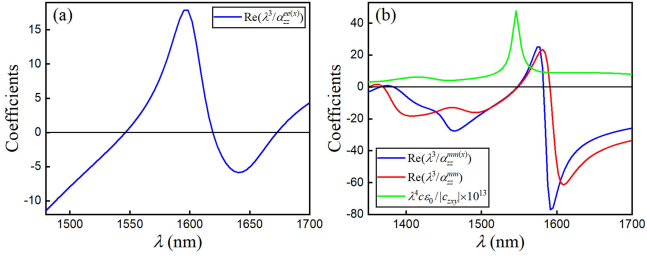}
\caption{Real (blue curves) and imaginary (red curve) parts of inverse (\textbf{a}) electric and (\textbf{b}) magnetic polarizabilities, as well as nonlocal response (green curve) for MoS$_2$ disk calculated using formulas (\ref{eq5}), (\ref{eq6}a) and (\ref{eq6}b), respectively. The simulation parameters: $R_2=H=418$~nm, $R_1=133$~nm, $\Delta_x=159$~nm.}\label{fig2}
\end{figure}

Note that the occurrence of a nonlocal response of the disk in Figure~\ref{fig2}b is associated with a cumulative effect, namely, the presence of a shifted hole, which makes NP asymmetrical, and is also the consequence of the anisotropy of the material. Formally, we can also include the nonlocal response in effective polarizabilities and use it in the further search for the conditions of QTM excitation:
\begin{equation}
\label{eq7}
\alpha_{zz}^{mm(y)}=\frac{m_z^{(E,k,-H)}-m_z^{(E,-k,H)}}{2H_z}.
\end{equation}
Figure~\ref{fig2}b shows a comparison of the frequency dependences for the inverse polarizabilities $\alpha_{zz}^{mm}$ and $\alpha_{zz}^{mm(y)}$. The difference between them is small, especially in the resonance region, where the inverse polarizability becomes zero. In contrast to the case of electric polarizability, there are only two such singular points, but their wavelengths are close to those obtained in Figure~\ref{fig2}a.

\section{Quasi-trapped modes of metasurfaces composed of MoS$_2$ disks}
\begin{figure*}[t]
\centering
\includegraphics[width=1.7\columnwidth]{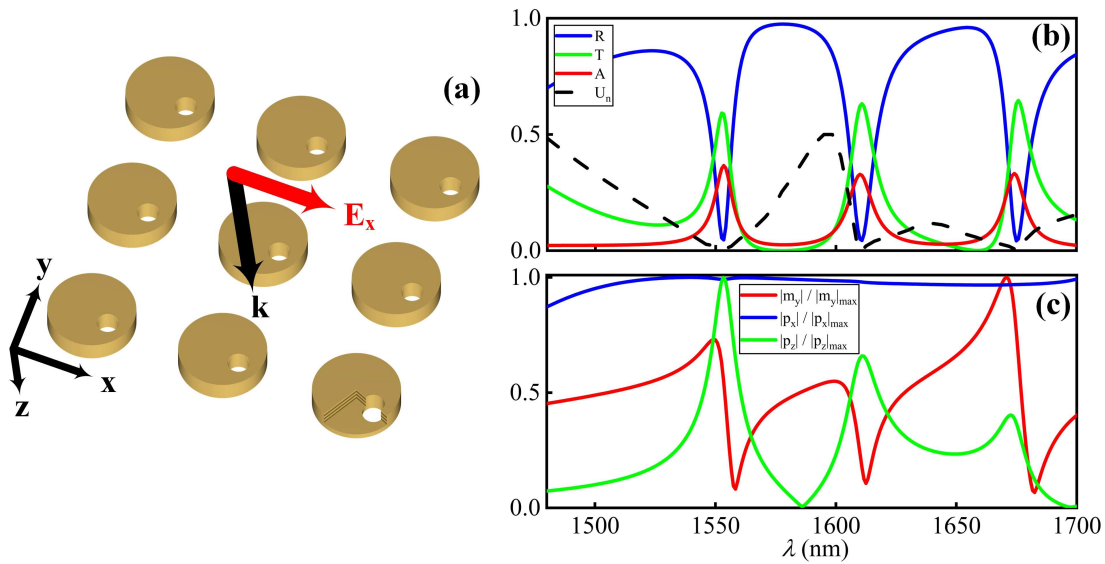}
\caption{(\textbf{a}) The model of metasurface based on MoS$_2$ disks with a hole shifted from the disk's center irradiated by normally incident wave $E_x(k_z)$. (\textbf{b}) Dependences of the reflection (R), transmission (T), and absorption (A) coefficients and parameter $U_n=0.5U/U_{max}$, as well as (\textbf{c}) the dependences of the $m_y$-component of the magnetic dipole and the $p_x$($p_z$)-component of the electric dipole excited in each disk on the wavelength for a metasurface composed of MoS$_2$ disks. The simulation parameters: $P=1125$~nm, $R_2=H=418$~nm, $R_1=133$~nm, $\Delta_x=159$~nm.}\label{fig3}
\end{figure*}

The components of electric or magnetic dipoles (\ref{eq3}) excited in each MoS$_2$ disk normally to the plane of its base can be used to excite the quasi-trapped modes in a metasurface composed of such disks~\cite{Evlyukhin2020}, see Figure~\ref{fig3}a. In this case, switching between $p_z$ and $m_z$ related to the electric and magnetic bianisotropy, respectively, can be simply performed by changing the polarization of the incident wave from $E_x$ to $E_y$.

In the case when optical losses are not taken into account, the formation of TM in an infinite periodic array can be realized due to the collective effect of the synchronization of $p_z$ or $m_z$ moments of individual disks. Assuming that the excitation of the corresponding dipole moments $p_z$ or $m_z$ is associated only with the direct polarizabilities $\alpha_{zz}^{ee}$ and $\alpha_{zz}^{mm}$, the condition for the existence of trapped electric and magnetic modes corresponds to the solution of transcendental equations written in the form~\cite{Babicheva}:
\begin{subequations}
\label{eq8}
\begin{align}
S_z^{(\textup{R})}&=\textup{Re}\bigg(\frac{1}{\alpha_{zz}^{ee(mm)}}\bigg),\\
S_z^{(\textup{I})}&=\textup{Im}\bigg(\frac{1}{\alpha_{zz}^{ee(mm)}}\bigg),
\end{align}
\end{subequations}
where $S_z=S_z^{(\textup{R})}+iS_z^{(\textup{I})}$ is the dipole sum whose components can be defined as follows:
\begin{subequations}
\label{eq9}
\begin{align}
S_z^{(\textup{R})}&=\frac{k_d^2}{4\pi}\sum_{l,j}^{\infty}\bigg(\frac{\cos(k_d d_{lj})}{d_{lj}}-\frac{\sin(k_d d_{lj})}{k_d d^2_{lj}}-\frac{\cos(k_d d_{lj})}{k_d^2 d^3_{lj}}\bigg),\\
\nonumber
S_z^{(\textup{I})}&=\frac{k_d^2}{4\pi}\sum_{l,j}^{\infty}\bigg(\frac{\sin(k_d d_{lj})}{d_{lj}}\\
&\mspace{80mu}+\frac{\cos(k_d d_{lj})}{k_d d^2_{lj}}-\frac{\sin(k_d d_{lj})}{k_d^2 d^3_{lj}}\bigg)\equiv-\frac{k_d^3}{6\pi}.
\end{align}
\end{subequations}
Here the parameter $d_{lj}=P\sqrt{l^2+j^2}$ is the distance from the Cartesian coordinate system origin, coinciding with the position of a lattice node, to all other nodes of the lattice, which are numbering by indices $l$ and $j$, $P$ is the period of lattice, $k_d=2\pi/\lambda_{QTM}$.

Condition (\ref{eq8}a) corresponds to the ideal situation of TM excitation without leakage; therefore, it allows one to determine the spectral position of quasi-trapped mode only approximately, while the deviation from condition (\ref{eq8}b) determines the decrease in the quality factor of the corresponding resonance. In the general case, the solution of equation (\ref{eq8}a) determines the exact value of the metasurface’s period $P$ corresponding to the QTM resonance at a given wavelength $\lambda_{QTM}$. On the other hand, the highest concentration of energy in a metasurface occurs when the resonance conditions are excited in each of its building blocks, which corresponds to the satisfaction of the condition $P/\lambda_{QTM}=0.72$~\cite{t16}. For the considered case, we obtain $P=1125$~nm and this value is well in agreement with the wavelength $\lambda_{QTM1}=1553$~nm for one of the resonances in Figure~\ref{fig3}b.

Note that the nonmonotonic behavior of the polarizability curve in the studied spectral region in Figure~\ref{fig2}a gives two more zeros of the function $U=\Big|S_z^{(\textup{R})}-\textup{Re}\big(\frac{1}{\alpha_{zz}^{ee}}\big)\Big|$ at the wavelengths 1619~nm and 1665~nm, which, together with $\lambda_{QTM1}$, determines the excitation of three QTM resonances at once in the considered spectral range. Indeed, the numerical simulation allows to observe these two additional resonances at wavelengths $\lambda_{QTM2}=1610$~nm and $\lambda_{QTM3}=1675$~nm, respectively. A feature of the resonances shown in Figure~\ref{fig3}b, in contrast to Ref.~\cite{ourPRB2022}, is the suppression of light reflection by the metasurface at QTM wavelengths. At the same time, the $p_x$ component for the individual disk from the metasurface remains almost at the same level in the whole spectral range in Figures~\ref{fig3}b and~\ref{fig3}c, while a resonant decrease in the $m_y$ component is observed through its bianisotropic coupling with $p_z$ determined by the term $\alpha_{zy}^{em}H_y$ in Equation~(\ref{eq4}a), see Figure~\ref{fig3}c. As a result, for the considered wavelengths $\lambda_{QTM}$, the energy of the wave incidenting on the metasurface is reallocated between the excitation of the QTM and the resonant amplification of the wave transmitted through the metasurface, which can be used for frequency selection. The important feature of QTM formation using strongly anisotropic NPs is the asymmetric field distribution inside the metasurface. This can be used to create near-field waveguide systems~\cite{Evlyukhin2006,Bozhevolnyi2006} in the metasurface.

\begin{figure*}[t]
\centering
\includegraphics[width=1.7\columnwidth]{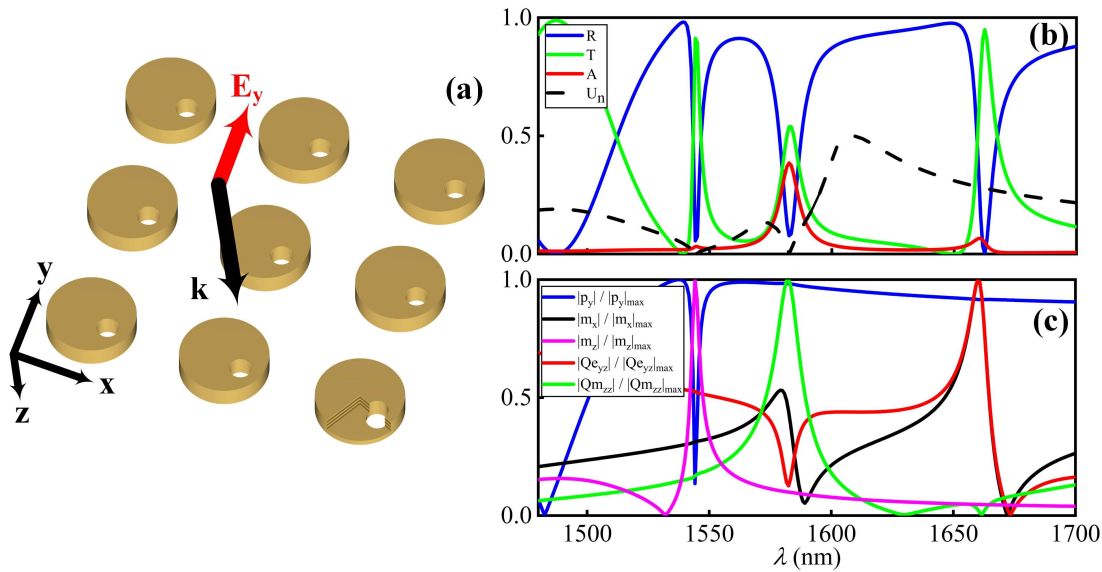}
\caption{(\textbf{a}) The model of metasurface based on MoS$_2$ disks with a hole shifted from the disk's center irradiated by normally incident wave $E_y(k_z)$. (\textbf{b}) Dependences of the reflection (R), transmission (T), and absorption (A) coefficients and parameter $U_n=0.5U/U_{max}$, as well as (\textbf{c}) dependences of various multipole components excited in each disk of metasurface on the wavelength. The parameters of metasurface correspond to Figure~\ref{fig3}.}\label{fig4}
\end{figure*}

\section{Polarization switching between different types of QTMs in the lattice of MoS$_2$ disks}
Changing the polarization of the incident wave to $E_y(k_z)$, see Figure~\ref{fig4}a, we again observe 3 resonances in the studied spectral range, but their wavelengths shift to the values $\lambda_{QTM1}^m=1545$~nm, $\lambda_{QTM2}^m~=~1583$~nm and $\lambda_{QTM3}^m=1663$~nm, see Figure~\ref{fig4}b. Analysis of the reasons for the appearance of these resonances shows that only for resonance at the wavelength $\lambda_{QTM1}^m$ the condition $U=\Big|S_z^{(\textup{R})}-\textup{Re}\big(\frac{1}{\alpha_{zz}^{mm}}\big)\Big|=0$ is satisfied, therefore, this resonance is associated with the trapped mode excitation. However, in the considered case, the QTM resonance is determined by the induced bianisotropy (see Equations~(\ref{eq4}b) and~(\ref{eq4}c)) of the magnetic type in each disk. The multipole analysis shown in Figure~\ref{fig4}c for the individual disk from the considered metasurface again demonstrates a similar mechanism for suppression of the reflection coefficient at wavelength $\lambda_{QTM1}^m$. In particular, the resonant amplification of the $m_z$ component of the magnetic dipole occurs in the QTM regime, which leads to the considerable decrease in the $p_y$ component of the disk due to the bianisotropic coupling through the term $\alpha_{zy}^{me}E_y$ in Equation~(\ref{eq4}b).

Note that the resonance at the wavelength $\lambda_{QTM3}^m$ is due to strong coupling between magnetic dipole and electric quadrupole, which corresponds to the simultaneous and sharp increase in intensity of $Qe_{yz}$  and $m_x$ in the vicinity of the wavelength 1660~nm~\cite{Babicheva2019}. Multiresonance for $\lambda_{QTM2}^m$ has a more complex nature and is due to the secondary excitation of several multipoles at once, including $Qm_{zz}$  component of magnetic quadrupole at the wavelength 1583~nm, see Figure~\ref{fig4}c.

\begin{figure*}[t]
\centering
\includegraphics[width=1.7\columnwidth]{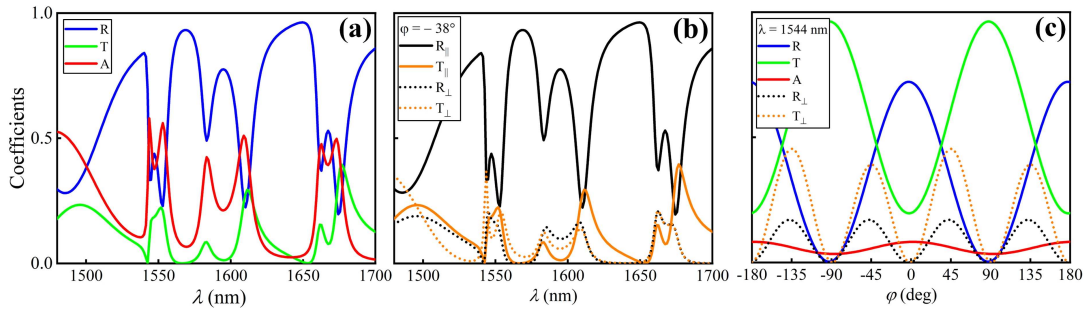}
\caption{(\textbf{a}) Dependences of reflection (R), transmission (T), and absorption (A) coefficients on the wavelength for the metasurface composed of MoS$_2$ disks irradiated by the wave with polarization oriented at angle $\varphi=-38^{\circ}$ relative to the $x$ axis. (\textbf{b}) Dependences of the reflection and transmission coefficients for parallel (R$_\|$, T$_\|$) and perpendicular (R$_\perp$, T$_\perp$) components of electric field on the wavelength for metasurface irradiated by wave with the polarization angle $\varphi=-38^{\circ}$. (\textbf{c}) Angular dependences of total (R, T, A) and orthogonal (R$_\perp$, T$_\perp$) reflection (R), transmission (T), and absorption (A) coefficients at the wavelength $\lambda=1544$~nm. The metasurface parameters correspond to the Figure~\ref{fig4}.}\label{fig5}
\end{figure*}

In practice, the presented triple resonance can be utilized for switching between the regimes of transmission/reflection of the metasurface at three wavelengths simultaneously. In particular, the change in the polarization of a normally incident wave from $E_x(k_z)$ to $E_y(k_z)$ leads to the switching of the reflection coefficient, for which the value of contrast can be defined as follows:
\begin{equation}
\label{eq10}
V=\frac{|\textup{R}_{E_x}-\textup{R}_{E_y}|}{\textup{R}_{E_x}+\textup{R}_{E_y}}\times100,
\end{equation}
where $R_{E_x(E_y)}$ corresponds to the reflection coefficients of the metasurface irradiated by the waves $E_x(k_z)$ or $E_y(k_z)$, respectively. In considered case, its values equal to: $V_1=68\%$ for $\lambda_{QTM1}=1553$~nm, $V_2=82\%$ for $\lambda_{QTM2}=1610$~nm, and $V_3=45\%$ for $\lambda_{QTM3}=1675$~nm.

Finally, we studied the peculiarities of QTM resonances under irradiation by normally incident waves at a certain angle $\varphi$ relative to the $x$ axis of the lattice, see Figure~\ref{fig5}a. Note that each feature degenerates into a double peak. This is due to the fact that each component ($E_x$, $E_y$) of the incident wave excites a certain type of resonance. Moreover, such an excitation of the metasurface leads to the effect of birefringence owing to the difference in the effective permittivities along the $x$ and $y$ axes. This is due to the combination of the asymmetry shape of MoS$_2$ disks with a hole and anisotropy of the material. As a result, a cross-polarization effect is observed for the considered metasurface accompanied by partial energy transfer from the incident wave into the orthogonal component of the reflected/transmitted wave, see Figure~\ref{fig5}b. In particular, the angular dependences of the corresponding coefficients R$_\perp$ and T$_\perp$ together with the total coefficients R, T, and A for the wavelength 1544~nm are shown in Figure~\ref{fig5}c. Thus, the metasurface composed of MoS$_2$ disks allows to perform the polarization control of the transmission/reflection regimes at several wavelengths at once, as well as to control the polarization characteristics of scattered light.

\section{Conclusion}
In present work, we have studied the frequency dependences of the resonant optical response of anisotropic MoS$_2$ nanoparticles possessing several bright resonances at once in near-infrared region. The scattering cross section has been calculated and multipole decomposition has been fulfilled for various directions of irradiation of MoS$_2$ disks with a hole. It has been shown that the presence of several multipole resonances of the same order at once in a narrow spectral range is associated with the internal anisotropy of the material, which also affects the nonlocal optical properties of the disks. Using the results of numerical calculation of the dipole polarizabilities for a single MoS$_2$ disk, the parameters of the metasurface, supporting several QTM resonances at once in the narrow spectral range of the near-infrared region, have been tuned. Each QTM resonance is related to the excitation of bianisothoric response of a single disk and characterized by a dip in the reflection coefficient and enhancement of light transmission by the entire metasurface in a narrow spectral range. The use of orthogonal polarization of the incident wave leads to the disappearance of the electric-type bianisotropic response, but causes the multiple bianisotropic magnetic-type responses responsible for the excitation of another multiresonance in the same spectral range. As a result, the switching of light reflection/transmission coefficient of the metasurface occurs, which can be used to design selective polarizing mirrors, optical switches, and routers with enhanced bandwidth. From a technical point of view, the designed metasurfaces can be fabricated by exfoliating individual flakes from transition metal dichalcogenide crystals and using their further processing by means of electron-beam lithography to form ordered arrays of nanoparticles.
\\

\section*{Author Contributions}
Conceptualization and methodology, A.V.P. and V.S.V.; formal analysis, A.V.P. and M.Y.G.; software and investigation, M.Y.G. and A.V.S.; visualization, M.Y.G. and A.V.S.; writing original draft, A.V.P., V.S.V. and M.Y.G.; writing—review and editing, A.V.P. and V.S.V.; funding acquisition, A.V.P.  and V.S.V.
All authors have read and agreed to the published version of the manuscript.

\section*{Conflicts of Interest}
The authors declare no conflict of interest

\begin{acknowledgments}
Authors thank A.B. Evlyukhin for helpful discussions. This research was funded by the Russian Science Foundation [grant number 22-22-01020].
\end{acknowledgments}

\bibliography{TripleArXiv}

\end{document}